\documentclass[aoas,MSNbibl,nameyear,dvips]{arximspdf}
\usepackage{graphicx}
\doi{10.1214/13-AOAS635} 
\volume{7}
\issue{3}
\pubyear{2013}
\firstpage{1663}
\lastpage{1683}

\makeatletter

\newtheorem{theorem}{Theorem}

\newcommand{\vX}{\mathbf{X}}
\newcommand{\vZ}{\mathbf{Z}}

\makeatother

\begin{document}
\begin{frontmatter}

\title{Learning local directed acyclic graphs based on multivariate
time series data\thanksref{T1}}
\runtitle{Local DAG on time series data}

\thankstext{T1}{Supported by NIH Grants R01CA127334 and R01GM097505.}

\begin{aug}
\author[A]{\fnms{Wanlu} \snm{Deng}\thanksref{T2,T3}\ead[label=e1]{wanlu.deng@pku.edu.cn}},
\author[A]{\fnms{Zhi} \snm{Geng}\thanksref{T3}\ead[label=e2]{zgeng@math.pku.edu.cn}}
\and
\author[B]{\fnms{Hongzhe} \snm{Li}\corref{}\ead[label=e3]{hongzhe@upenn.edu}}
\runauthor{W. Deng, Z. Geng and H. Li}
\affiliation{Peking University, Peking University and University of
Pennsylvania}
\address[A]{W. Deng\\
Z. Geng\\
Department of Statistics\\
\quad and Probability\\
Peking University\\
Beijing 100871\\
China\\
\printead{e1}\\
\hphantom{E-mail: }\printead*{e2}}
\address[B]{H. Li\\
Department of Biostatistics\\
University of Pennsylvania\\
School of Medicine\\
Philadelphia, Pennsylvania 19104\\
USA\\
\printead{e3}} 
\end{aug}

\thankstext{T2}{Supported in part by the China State Scholarship Fund.}

\thankstext{T3}{Supported by the National Natural Science Foundation of China
(11171365, 10931002, 11021463) and the Center for Statistical Science,
Peking University.}

\received{\smonth{9} \syear{2011}}
\revised{\smonth{2} \syear{2013}}

%
\begin{abstract}
Multivariate time series (MTS) data such as time course gene
expression data in genomics are often collected to study the dynamic
nature of the systems. These data provide important information about
the causal dependency among a set of random variables. In this paper,
we introduce a computationally efficient algorithm to learn directed
acyclic graphs (DAGs) based on MTS data, focusing on learning the local
structure of a given target variable. Our algorithm is based on
learning all parents (P), all children (C) and some descendants (D)
(PCD) iteratively, utilizing the time order of the variables to orient
the edges. This time series PCD-PCD algorithm (tsPCD-PCD) extends the
previous PCD-PCD algorithm to dependent observations and utilizes
composite likelihood ratio tests (CLRTs) for testing the conditional
independence. We present the asymptotic distribution of the CLRT
statistic and show that the tsPCD-PCD is guaranteed to recover the
true DAG structure when the faithfulness condition holds and the tests
correctly reject the null hypotheses. Simulation studies show that the
CLRTs are valid and perform well even when the sample sizes are small.
In addition, the tsPCD-PCD algorithm outperforms the PCD-PCD algorithm
in recovering the local graph structures. We illustrate the algorithm
by analyzing a time course gene expression data related to mouse
T-cell activation.
\end{abstract}

%

\end{frontmatter}

\section{Introduction}
Inferring causal networks among a set of genes based on their
expression levels is one of the most important problems in genomics.
High-throughput technologies such as microarrays or next generation
sequencing have enabled biologists to measure expression levels of all
the genes in large-scale. Technologies are also available to obtain
gene expressions at single-cell level for inference of single-cell
expression dynamics. Time-course gene expression experiments, where the
expression levels of the genes are measured over time during a
biological process, are particularly important in providing dynamic
information about gene regulation and networks [\citet{Buganim}]. The
focus of this paper is on learning the local graphs based on such
multivariate time course gene expression data.

Graphical models have been applied to study gene networks based on gene
expression data, among which Gaussian graphical models are most
commonly used and studied [\citet{Strimmer1,LiGui,PeiWang12}].
However, the Gaussian graphical models only provide information on
conditional independence and the resulting graphs are undirected.
Alternatively, directed acyclic graphs (DAGs) are frequently used to represent
independence, conditional independence and causal relationships among random
variables in a complex system [\citet{Pearl2000,Spirtes2000}]. In
such DAG models, parents of some nodes in the graph are
understood as ``causes,'' and edges have the meaning of ``causal
influences.'' The
causal influences among random variables imply conditional independence
relations
among them. \citet{Murphy1999} and \citet{Friedman2000} have suggested
using Bayesian network models of gene expression networks.

Methods for learning the structures of
DAGs include the search-and-score based methods [\citet
{Cooper1992,Heckerman1995,Chickering2002,Friedman2003}] that require
elicitation of all
the conditional probabilities and the constraint-based methods
[\citet
{Neapolitan2003}]
that evaluate the presence or
absence of an edge by testing conditional independence among variables.
These constraint-based learning methods often
require unreasonable amounts of data in order to accurately estimate
higher order conditional independence relations from finite samples.
Algorithms that combine ideas from constraint-based and
search-and-score techniques have also been developed and have shown
excellent performance [\citet{Tsamardinos2006}].
Efficient Markov chain Monte Carlo methods have also been developed for
learning the Bayesian networks [\citet{Friedman2003,Ellis2008}].

Most of the current available methods for structural learning of DAGs
assume that the data are i.i.d. samples from some joint distribution
specified by
the underlying DAG. These methods cannot be directly applied to
multivariate time series (MTS) data.
One approach to causal graph learning from the MTS data is based
on the dynamic Bayesian network (DBN) model [\citet{Ghahramani1997}],
which is an extension of
the Bayesian network model for time series data. DBN models the
stochastic evolution of a set of random variables over time. In
comparison with the Bayesian network, discrete time is introduced and
conditional distributions are related to the values of parent variables
in the previous time point. Moreover, in DBNs the acyclicity constraint
is relaxed. \citet{Husmeier2003} studied the sensitivity and
specificity of inferring genetic
regulatory interactions from microarray
experiments with DBNs. \citet{Husmeier2011} presented methods for
improvements in the reconstruction of time-varying gene regulatory
networks using dynamic programming and regularization by information
sharing among genes. \citet{Rau2010} proposed to apply the linear
Gaussian state-space models, a subclass of DBNs, for estimating
biological networks from time course gene expression data with replications.

The DAG structural learning algorithms [\citet
{Pearl2000,Spirtes2000,Heckerman1995,Tsamardinos2006}] have mainly
focused on constructing
the whole directed graph over
all the variables. Such whole directed graphs are often difficult to
learn due to small sample sizes or
limited perturbation to the underlying system to infer the causal
relationships. In
some applications, one might be interested in identifying the causal
variables of a
given node and the variables that this node influences, that is, in
identifying the
local structure of a variable. In genomics, we might be interested in
learning the
upstream regulators of a gene and also the downstream genes regulated
by this gene.
Methods for learning the local directed graphs have many practical
applications and play a central role
in causal discovery and classification because of their scalability
benefits. One key concept in learning
local causal graph structure is the Markov blanket of a variable $T$,
which is a minimal variable subset conditioning on which all
other variables are probabilistically independent of $T$. Finding the
Markov blanket has been the basis of many of the newly developed
methods for learning DAG structures [\citet
{Margaritis2000,Tsamardinos2003,Tsamardinos2006}].

\citet{Yin2008} developed a partial
orientation and local structure learning algorithm based on identifying
all parents
(P), all children (C) and some descendent (D) of a given node. This
algorithm is
called the PCD-PCD algorithm. \citet{Zhou2010} further extended
this algorithm
by constructing a larger local network with depth $d$ and using a more efficient
stopping rule to reduce the computation time. Like many other
structural learning
algorithms, PCD-PCD is based on the tests of conditional
independence among a set of variables. \citet{Yin2008,Zhou2010} used
the standard likelihood ratio tests (LRTs) for conditional independence
testing. However,
such tests cannot be applied directly to MTS data due to the dependency
of the data observed over time.


In this paper, we propose to extend PCD-PCD to MTS data in order to
learn a local network around a target variable. We consider
stationary ergodic MTS with time-invariant dependence
structure.
In our approach, the search for separators of a
pair of variables in a large DAG is
localized to small subsets and, thus, the approach can improve the
efficiency of
searches and the power of statistical tests for structural learning.
Our approach
captures the Bayesian dynamic nature of the dependence by learning the
structure of
the graphical model based
on conditional independence between the past and future of observations
of the time
series. The new method utilizes the time order to orient edges
connecting variables
at different time points. Like the PCD-PCD algorithm, we first find
parents, children and some
descendants of the target $T$ to obtain a local skeleton with a radius
1, and then
repeatedly find PCDs of the nodes in the previous PCDs until the radius
of the local
skeleton is up to the given depth $d$. By focusing on learning the
local DAGs, the proposed algorithm
can handle high-dimensional random variables even when the sample size
is not too large.

Since the conditional independence test plays a key role in our
proposed algorithm, we develop composite likelihood ratio tests
(CLRTs) for conditional independence for MTS data, taking into account the
dependency of the variables over time. The commonly used likelihood
ratio test for conditional
independence is invalid under the small
sample size when the null hypothesis only involves variables at the
same time point.
The CLRT statistic enables us to perform valid inference on conditional
independency
under the setting of small sample sizes, allowing the number of
independent time
series smaller than the number of variables at each time point.

In Section \ref{algo} we present the local structural learning algorithm
tsPCD-PCD for multivariate time series data. In Section \ref
{complike} we develop the CLRTs for conditional
independence for time series data. We then present simulation results
to evaluate
the algorithm and the validity of the CLRTs in Section \ref{simu} and
application of the
method to analysis of gene expression data related to T-cell activation
in Section~\ref{real}. Finally, we give a brief discussion of the methods in
Section \ref{disc}. The exact statements of the theorems and their
proofs can be found in the supplementary material [\citet{DenGenLi13}].

\section{Algorithm for learning a local structure around a target
variable}\label{algo}
\subsection{Statistical model, data observed and notation}
We consider the data from MTS where the lag of the time series is
$q$ and there are $p$ variables at each time point. Let
$\vX_t=(X_{t,1},\ldots,X_{t,p})^{\prime}$ denote the $p$-dimensional
random vector at time $t$, for $t=1,2,\ldots, n$. We assume that
$\{\vX_t, t=1,\ldots,n \}$ is stationary and ergodic. In addition,
we assume that the dependency structure of $\{\vX_t, t=1,\ldots,n
\}$ is determined by a DAG $\mathcal{G}$ with time-order
constraint. Here a DAG is defined as $\mathcal{G} = (A,E)$, while $A$
is a
finite set of nodes, and $E$ is a set of directed edges on~$A$, with
no directed cycle. These directed edges in $\mathcal{G}$ may involve
variables at the same time points or between different time points.
We assume that the true DAG is time-invariant. Based on the
stationary assumption, to recover $\mathcal{G}$, we only need to
learn the directed edges among the variables $\vX_t$ and the edges
connecting variables in $\{ \vX_{t-l}, 1 \leq l \leq q \}$ and
variables in $\vX_t$ for a given time point $t$. In addition, we
assume that if there is a link between $X_{tg}$ and $X_{t-l,g'}$,
then $X_{t-l,g'}$ causes $X_{tg}$. Using the language of a graphical
model, we say there is a direct edge between $X_{t-l,g'}$ and
$X_{tg}$, that is, $X_{t-l,g'}\rightarrow X_{tg}$.

Instead of learning the DAG on the variables $(\vX_{t-q}',\ldots,
\vX_{t-1}', \vX_{t}')'$, which is of $p\times(q+1)$ dimension, we
are interested in learning the variables with direct edges to and
from a target variable $T$. For the observed data, suppose there are
$m$ i.i.d. MTS data, each of which is a time series with length of
$n_j$, $j=1,\ldots,m$, and with the same time lag $q$. Let
$\vX_{t}^{(j)}= (X_{t,1}^{(j)},\ldots,X_{t,p}^{(j)})'$ be the vector
of $p$ variables for the $j$th sequence at time $t$ for
$t=1,\ldots,n_j$ and $j=1,\ldots,m$. We can rewrite the observed
data
by piling them up according to the lag $q$. Specifically, for the
$j$th time series, we rewrite the data $(\vX_1^{\prime(j)}, \vX_2^{\prime(j)},\ldots,
\vX_{n_j}^{\prime(j)})'$ as
\[
\matrix{ X_{1,1}^{(j)} & \cdots& X_{1,p}^{(j)}
& X_{2,1}^{(j)} & \cdots& & \cdots& X_{q+1,j}^{(j)}
& \cdots& X_{q+1,p}^{(j)},
\vspace*{2pt}\cr
X_{2,1}^{(j)} & \cdots&X_{2,p}^{(j)} &
X_{3,1}^{(j)} & \cdots& & \cdots& X_{q+2,1}^{(j)}
& \cdots& X_{q+2,p}^{(j)},
\cr
& & & & & \vdots& & & &
\cr
X_{n_j-q,1}^{(j)} & \cdots& X_{n_j-q,p}^{(j)} &
X_{n_j-q+1,1}^{(j)} & \cdots& & \cdots& X_{n_j,1}^{(j)}
& \cdots& X_{n_j,p}^{(j)}.}
\]
Because the series is stationary and ergodic, each row of this
piled data set has $p\times(q+1)$ variables and all rows have the
same joint distribution. However, these rows are not independent and
piled data include $N=\sum_{j=1}^{m} n_j - qm$ dependent
observations. Let $T$ be the target variable, which is one of the
$p$ variables. Due to stationarity and without loss of generality,
we view $T$ as a node at the time point~$t$, and let $A$ denote the
full set of nodes in the set $A=(\vX_{t-q}', \ldots, \vX_{t-1}',
\vX_t')'$. Our goal is to identify the nodes in $A$ that are linked
to $T$ based on the piled data. Since we assume that the time
series are stationary, the local structure around node $T$ is
time-independent, which enables us to utilize the piled data to
learn the local structure.

Before we introduce our proposed algorithm, we give some definition and
notation.
We say that the probability distribution $P$ and the DAG $\mathcal{G}$
are connected by the Markov condition property if a node is
conditionally independent of its nondescendants given its parents.
A DAG $\mathcal{G}$ and a joint distribution $P$ are faithful to one
another, if every conditional independence entailed by the graph of
$\mathcal{G}$ and the Markov condition is also present in $P$
[\citet{Spirtes2000}]. For a node $u$, let PC[$u$] denote the set of
all parents and all children of $u$, Pa[$u$] denote the parent
nodes, Ch[$u$] denote the children, and let PCD[$u$] denote a set
that contains PC[$u$] and may contain some descendants of $u$. For a
subset $B\subset A$ of the vertices of $\mathcal{G}$, the induced
subgraph on $B$ is $\mathcal{G}[B]:= (B,E[B])$, where $E[B]:=
E\cap(B\times B)$. A $v$-structure [also called immorality by, e.g.,
\citet{Lauritzen1996}] is an induced subgraph of $G$ of the form
$a \rightarrow b \leftarrow c$. The existence of a $v$-structure
among a set of three variables can be determined by conditional
independence tests [\citet{Lauritzen1996}]. The skeleton of a DAG
$\mathcal{G}$ is the undirected graph $\mathcal{G}^u:= (A,E^u),
E^u:= \{(a, b) \in A \times A | a\rightarrow b$ or $b\rightarrow
a \in\mathcal{G}\}$. Two DAGs are called Markov equivalent if they
induce the same conditional independence restrictions. Two DAGs are
Markov equivalent if and only if they have the same global skeleton
and the same set of v-structures [\citet{Verma1990}]. An equivalence
class of DAGs consists of all DAGs that are
Markov equivalent. Finally, throughout this paper, we use lower
case single letters to present the nodes of the graph unless
otherwise specified.

%
%

\subsection{A brief review of the max--min parents and children algorithm}
Our algorithm depends on finding the PCD of a given node $u$, PCD[$u$].
The max--min parents and children (MMPC) algorithm originally proposed in
\citet{Tsamardinos2003} presents a computationally feasible algorithm
to find PCD[$u$] for a given node $u$ and can be efficiently applied to
thousands of variables. The algorithm was further studied and justified
by \citet{Tsamardinos2006}.
MMPC run on target node $u$ provides a way to identify the existence of
edges to and from $u$
(but without being able to identify the orientation of the edges).
It is a two-phase
algorithm. In phase I, the forward phase, variables enter the
candidate set of PCD[$u$] sequentially, called CPCD by using the
max--min heuristic: in each iteration, select the variable that
maximizes the minimum association with $u$ relative to CPCD, and add
it to CPCD; the iteration stops when the maximum is zero, that is, all
remaining variables are independent of the target $u$ given some
subset of CPCD. Here the minimum association between $u$ and $v$ is
the minimum of association between these two variables achieved over
all subsets of CPCD. In phase II, the backward phase, the false
positives in CPCD, which are independent of $u$ given a subset of
CPCD, are removed from CPCD. This is achieved by testing the
conditional independence
between $u$ and $v$ given some subset of CPCD; if the null is not
rejected, $v$ is removed from CPCD.

\subsection{Algorithm for learning local structure around a target
$T$}

We extend the PCD-PCD algorithm of \citet{Zhou2010} for local
directed graph learning to MTS data by assuming that if there is a
link between the variables at different time points, the causal
direction is determined and is from the variable measured at the
early time point to the variable measured at the later time point.
Like the PCD-PCD algorithm, we first find parents, children and
some descendants of the target $T$ to obtain a local skeleton with a
radius of 1, and then repeatedly find PCDs of nodes that are in the
previous PCDs and also on the same time point as the target node
$T$, until the radius of the local skeleton within time point $t$ is
up to the given depth $d$. In order to orient the edges in the local
skeleton, sometimes it is necessary to find the PCDs further away
from the target along some but not all the paths. Note that some of
the undirected edges cannot be oriented from the observational data
due to the existence of equivalent class [\citet{Andersson1997}].
\citet
{Zhou2010} proposed
a stopping rule so that the process of finding PCDs can stop early
even if some edges within the local graph are not oriented. The
stopping rule is based on the fact that when the unoriented edges are
surrounded by directed edges, they cannot be oriented by finding
further structures.

Our algorithm is divided into two parts. Part I involves finding
the local structure around the target $T$ to the depth of $d-1$ and
part II finds the edges at the last layer $d$ and orients undirected
edges within the local structure with depth $d$. In both parts of
the algorithm, we use the index set $1\dvtx p(q+1)$ to denote the node
set $(\vX_{t-q}', \vX_{t-1}', \ldots, \vX_t')'$, which is ordered
according to the time order. Based on this notation, the nodes in
$\vX_t'$ are denoted as $(pq+1)\dvtx p(q+1)$ and the target $T$ is coded
with a number in $(pq+1)\dvtx p(q+1)$. Furthermore, we use $V$ to
denote a set of variables whose PCDs have been
obtained, $L[i]$ to denote the node set on the $i$th layer of the local
graph, $L$ to denote
the set of nodes in all layers. In addition, we use $C[i]$ to denote an
ordinal waiting list for layer $i$ whose PCD
is to be determined and $C$ to denote all the nodes at the current time
point. Finally, let
$D$ be the counter of the depth of the graph.

Part I of the algorithm is detailed in Table \ref{part1}. Part I stops
if all
nodes with a path to $T$ have a distance shorter than $d$ or the first $d-1$
layers of nodes around $T$ have been obtained. A detailed explanation
of the steps of this part of the
algorithm can be found as part of the proof of Theorem \ref{thm1}
presented below.

%
\begin{table}
\tablewidth=293pt
\caption{tsPCD-PCD algorithm, part I: find edges within depth $d-1$
from the
target node $T$}\label{part1}
%
%
\begin{tabular*}{\tablewidth}{@{\extracolsep{\fill}}ll@{}}
\hline
1 & \textbf{Initialization}: Find the PCD of $T$, PCD[$T$]. \\
& $V=\{T\}$, \\
& $L[0]=\{T\}$, $L(i)=\varnothing$ for $i=1,\ldots,d$, \\
& $L=\{T\}$, \\
& $D=1$, \\
& $C[0]=\{T\}$,\\
& $C[1]=\operatorname{PCD}[T]$,\\
& $C=(p*q+1)\dvtx p*(q+1)$, \\
& create directed edges $(\operatorname{PCD}[T] \setminus C \rightarrow T)$,
\\
& $C(1)=\operatorname{PCD}[T] \cap C$. \\
& \textbf{Repeat} \\
2 & Remove $x$ from the head of list $C[D]$. \\
3 & If $x$ $\notin V$, then \\
& \ \ \ Find PCD[$x$], and set $V=V \cup{x}$. \\
& \ \ \ create directed edges $(\operatorname{PCD}[x] \setminus C \rightarrow
x)$ \\
& \ \ \ $\operatorname{PCD}[x]=\operatorname{PCD}[x]\cap C$ \\
& \ \ \ For each $y \in V$, if $\{x \in\operatorname{PCD}[y]$ and $y \in
\operatorname{PCD}[x]\}$,\\
& \ \ \ \ \ \ then create an undirected edge $(x,y)$.\\
& \ \ \ Find $v$-structures for the triple of $x$, one of Pa$[x]$ on
previous \\
& \ \ \ \ \ \ time points and one of nodes on current time point that have
\\
& \ \ \ \ \ \ undirected edges with $x$, if $x$ is not in the separator
set\\
& \ \ \ \ \ \ of the last two nodes. \\

& \ \ \ Find $v$-structures within $V$ including $x$:\\
& \ \ \ \ \ \ \{Within $V$, find possible $v$-structures only for the
triple of $x$
\\
& \ \ \ \ \ \ and other two variables in $V$ if an intermediate node is not
\\
& \ \ \ \ \ \ in the separator set of two nonadjacent nodes.\}\\
& \ \ \ Orient undirected edges under Meek's rules [\citet
{Meek1995}]:\\
& \ \ \ \ \ \ \{Orient other edges between nodes in $V$ if each
opposite of \\
& \ \ \ \ \ \ them creates either a directed cycle or a new
$v$-structure.\}\\
& End if \\
4 & If $x$ $\notin$ $L$ and $x$ $\notin$ $L[D]$
and
$x$ is adjacent to a node in $L[D-1]$ then \\
&\ \ $L[D]=L[D] \cup\{x\}$ and add
$(\operatorname{PCD}[x] \cap C) \setminus L$ to the tail of
list\\
& \ \  $C[D+1]$. \\
& End if. \\
5 & If $C[D]=\varnothing$ then \\
& \ \ \ $L= L \cup L[D]$ and $D
=D+1$ \\
& End if. \\
6 & \textbf{Until} $C[D] = \varnothing$ or $D \geq d$.\\
\hline
\end{tabular*}
\end{table}

When $d>1$, part II of the algorithm is required to identify the edges
at the last layer $d$ and to orient the undirected edges within the
local structure with
depth $d$. In part II, we use the notation
\[
\mbox{\textit{struct}}(\mbox{``\textit{leaf},'' } v, \mbox{``\textit{length},'' } l,
\mbox{``\textit{path},'' } u)
\]
to define a set with three different elements, ``leaf'' ($v$),
``length''
($l$) and ``path''~($u$),
where a ``leaf'' is a node $v$ at layer $\ge d$, ``path'' $u$ is a set of
variables on the path from layer $d-1$ to the ``leaf'' $v$
and ``length'' is the length from layer $d-1$ to ``leaf''~$v$. For a given element $x$ in this set, $x.\mbox{\textit{leaf}}, x.\mbox{\textit{path}}$ and
$x.\mbox{\textit{length}}$ denote
the three elements of the list, respectively.
Details of part II of the tsPCD-PCD algorithm are
given in Table \ref{part2}. A~detailed explanation of the steps of
part II can be found as part of the proof of Theorem~\ref{thm1}.

%
\begin{table}
\tablewidth=292pt
\caption{tsPCD-PCD algorithm, part II: find edges at layer $d$ and orient
undirected edges within the local structure}\label{part2}
\begin{tabular*}{\tablewidth}{@{\extracolsep{\fill}}ll@{}}
\hline
1 & \textbf{Initialization}: Learn the PCDs of nodes at the layer $d-1$,
\\
&and construct a set $W$: \\
& $W = \{\mbox{\textit{struct}}(\mbox{``\textit{leaf},'' } v, \mbox{``\textit{length},'' } 1,
\mbox{``\textit{path},'' }
u)\dvtx u \in L[d-1]$, \\
& $v \in(\operatorname{PCD}[u] \cap C) \setminus L\}$ \\
& \textbf{Repeat} \\
2 & Remove $x$ from the head of list $W$. \\
3 & If all edges on path $x.\mbox{\textit{path}}$ are undirected then \\
& \ \ \ If $x.\mbox{\textit{leaf}}$ $\notin$ $V$ then \\
& \ \ \ \ \ \ Find PCD[$x.\mbox{\textit{leaf}}$] and set $V=V \cup\{x.\mbox{\textit{leaf}}\}$. \\
& \ \ \ \ \ \ Create directed edges $(\operatorname{PCD}[x.\mbox{\textit{leaf}}] \setminus C
\rightarrow x.\mbox{\textit{leaf}})$ \\
& \ \ \ \ \ \ $\operatorname{PCD}[x.\mbox{\textit{leaf}}]=\operatorname{PCD}[x.\mbox{\textit{leaf}}] \cap C$ \\
& \ \ \ \ \ \ For each $y \in V$, if $\{x.\mbox{\textit{leaf}} \in\operatorname{PCD}(Y)$ and
$y \in
\operatorname{PCD}[x.\mbox{\textit{leaf}}]\}$,\\
& \ \ \ \ \ \ \  \ create an undirected edge $(x.\mbox{\textit{leaf}},y)$.\\
& \ \ \ \ \ \ Find $v$-structures for the triple of $x.\mbox{\textit{leaf}}$, one of
Pa$[x.\mbox{\textit{leaf}}]$
\\
& \ \ \ \ \ \ \ \ on previous time points and one of nodes on current time
\\
& \ \ \ \ \ \ \ \
point that have undirected edges with $x.\mbox{\textit{leaf}}$, if $x.\mbox{\textit{leaf}}$
\\
& \ \ \ \ \ \ \ \  is not in the separator set of the last two nodes. \\
& \ \ \ \ \ \ Find $v$-structures within $V$ including $x.\mbox{\textit{leaf}}$.\\
& \ \ \ \ \ \ Orient undirected edges under the Meek's rules
[\citet
{Meek1995}].\\
& \ \ \ End if. \\
& \ \ \ If there is an undirect edge between ${x.\mbox{\textit{leaf}}}$ and the last
node $u$
\\
& \ \ \ \ \ \ of ${x.\mbox{\textit{path}}}$, then add \\
& \ \ \ \ \ \ $\{ \mbox{\textit{struct}}(\mbox{``\textit{leaf},'' }
v, \mbox{``\textit{length},'' } x.\mbox{\textit{length}}+1,
\mbox{``\textit{path},'' } [x.\mbox{\textit{path}},x.\mbox{\textit{leaf}}])$:\\
& \ \ \ \ \ \ $v \in \operatorname{PCD}[{x.\mbox{\textit{leaf}}}]
\cap{C}\setminus{x.\mbox{\textit{path}}}\setminus{L} \}$ \\
& \ \ \ \ \ \ to the tail of $W$. \\
& \ \ \ End if. \\
& End if. \\
4 & \textbf{Until} $W = \varnothing$. \\
& Return \\
\hline
\end{tabular*}
\end{table}

The following theorem shows the effectiveness of the algorithm for
recovering the true directed local graph structure for the Markov
equivalence class
of the underlying global
DAG.

%
\begin{theorem} \label{thm1}
Suppose that a DAG is faithful to the probability distribution $P$ of the
multivariate time series $\vX_t$ and all conditional independencies
can be correctly
inferred based on the data. Then for a given target node $T$, the tsPCD-PCD
algorithm can correctly recover the edges within a depth $d$ of the
local directed
graphical structure around $T$ at the same time point and the edges
connecting $T$
and the variables on previous time points. Furthermore, the algorithm
obtains the
same orientations of these edges as a partially directed graph for the Markov
equivalence class of the underlying global DAG.
\end{theorem}
The proof of this theorem, which is based on detailed explanations of
the steps of the
algorithm, is given in the supplementary material [\citet{DenGenLi13}].

%
\begin{figure}

\includegraphics{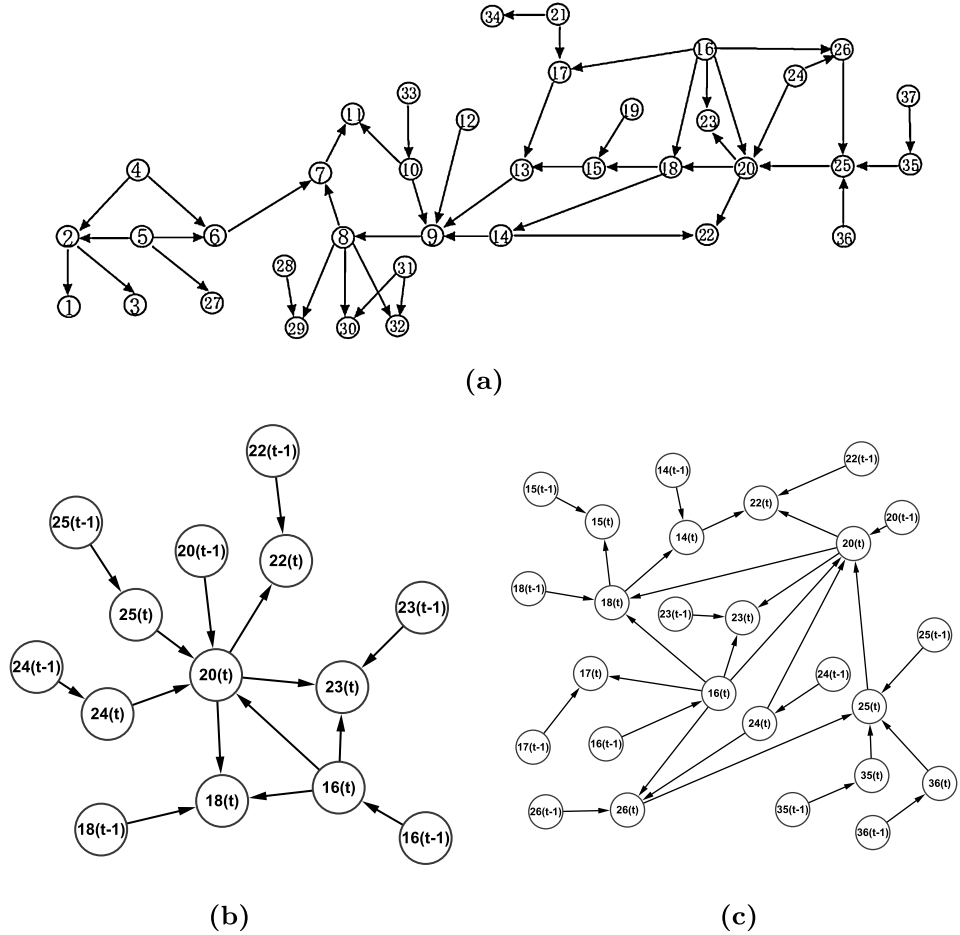}

\caption{\textup{(a)} The DAG used in simulations for variables on the same
time point; \textup{(b)} local network around node 20 learned by part I
of the
algorithm; \textup{(c)} local network around node 20 learned by part~II of
the algorithm.} \label{Fig4}
\end{figure}

\subsection{An illustrative example}\label{example}
As a simple illustration of the algorithm, we consider the ALARM
network structure as shown in Figure \ref{Fig4}(a)
[\citet{Beinlich1989}] with 37 nodes and 46 edges. This ALARM DAG has
been extensively used in evaluating DAG learning algorithms. We
extend this network to a dynamic DAG depicting stationary time
series with lag $q=1$ by assuming that each variable is also
influenced by itself on the previous time point. We aim to identify
the local network of node 20 to depth $d=2$. We choose node 20
since it has the largest degrees among all the nodes. Suppose that
the data are large enough to correctly identify all the required
conditional independence.
Figure \ref{Fig4}(b) shows a
a local network with $d=1$ that is determined by part I of the
algorithm and
Figure \ref{Fig4}(c) shows the final local network with
$d=2$ after applying part II of the algorithm. Note that there is no
guarantee that all edges can be oriented by the algorithm. If there
are some unoriented edges, we need to extend the network along the
undirected paths in order to orient those undirected edges in part
II. In this example, since all edges are oriented after learning the
first two layers, the algorithm stops extending to higher layers.

To illustrate part II, in the initialization step, we construct the
set $W$, which contains the potential undirected edges that may
need to be extended to higher layers for edge orientation within the
first $d=2$ layers. For a given node $x_{t,18} \in L[1]$, we have
PCD[$x_{t,18}] = \{x_{t-1,18}, x_{t,14}, x_{t,15}, x_{t,16},
x_{t,20}\}$. Since $x_{t-1,18} \rightarrow x_{t,18}$ and
$\{x_{t,20}, x_{t,16}\} \in L[0] \cup L[1]$, these three nodes
are not used as leafs in constructing the set $W$. We then
add
\[
\mbox{\textit{struct}}(\mbox{``\textit{leaf},'' } x_{t,14},
\mbox{``\textit{length},'' } 1, \mbox{``\textit{path},'' } x_{t,18})
\]
and
\[
\mbox{\textit{struct}}(\mbox{``\textit{leaf},'' } x_{t,15}, \mbox{``\textit{length},'' }
1, \mbox{``\textit{path},'' } x_{t,18}\}
\]
to $W$ for
node $x_{t,18}\in L[1]$. Similarly, we add other elements to $W$ for
each of the other nodes in $L[1]$ and finish the initialization of
$W$.

\section{Composite likelihood ratio tests for conditional
independence}\label{complike}
The tsPCD-PCD algorithm and also the validity of Theorem \ref{thm1}
depend on a valid and powerful test
for conditional dependency.
Specifically,
finding the PCD of a node using the MMPC algorithm and finding the
$v$-structures among a set of three variables both rely on testing for
conditional independence. For the time series data, since
the data across
different time points are dependent, the commonly used LRTs
tests cannot be applied directly. We propose to
develop composite likelihood ratio tests for conditional independence
and derive
their asymptotic distributions.

\subsection{The composite likelihood ratio tests for general
parametric models}
Consider a model for the joint density of $\vX=(\vX_{t-q}',\ldots,
\vX_t')$, denoted by
$P(\vX_{t-q}',\ldots, \vX_t'; \theta)$, where $\theta\in
\mathcal{H} \subseteq\mathbb{R}^k$ is a $k$-dimensional model
parameter, where $\mathcal{H}$ is an open set. The conditional
independence test is used in our algorithm for determining the PCDs
and $v$-structures. Depending on the models assumed, the null model
under the conditional independence assumption corresponds to
certain constraints on the model parameter $\theta$, that is,
$h(\theta)=0$ for some multidimensional function $h$ (see Lemma 2
in the supplementary material [\citet{DenGenLi13}]). If the
first-order partial derivative of $h$ is
continuous, it can be equivalently expressed as $\theta=
g(\varphi)$, where $\varphi$ is a parameter vector with dimension
lower than $k$. As an example, consider a joint multivariate model
with four variables
$X_1, X_2, X_3$ and $X_4$, with a joint density $\mathcal
{N}(0,\Sigma)$, where $\Sigma= (\sigma_{ij})$ is a $4\times4$
matrix. Under this simple model,
\[
H_0\dvtx X_1 \perp X_2 | X_3
\quad\Leftrightarrow\quad H_0\dvtx  \Sigma_{X_1,X_2,X_3}^{-1} (1,2) = 0
\quad\Leftrightarrow\quad H_0\dvtx \sigma_{12} = \frac{\sigma_{13}*\sigma_{23}}{\sigma_{33}}.
\]
Similarly, for categorical variables and log-linear models, the null
hypothesis of conditional independence corresponds to constraints on
model parameters [\citet{Lauritzen1996}].
We therefore consider the conditional independence null hypothesis that
can be expressed as
%
%
\begin{equation}
\label{h0} H_0\dvtx  \theta= g(\varphi),
\end{equation}
where $\theta\in\mathbb{R}^k$, $\varphi\in
\mathbb{R}^{k-r}$ and $g\dvtx\mathbb{R}^{k-r} \rightarrow
\mathbb{R}^{k}$ is a function with $\frac{\partial}{\partial
\varphi} g(\varphi)$ being of full rank. Under the conditional
independence null hypothesis, the parameter can be written as
$\mathcal{H}_0 =\{\theta\dvtx \theta= g(\varphi)\}$.

For multiple time series $\{\vX_{t}^{(j)}, t=1,\ldots,n_j,
j=1,\ldots,m\}$, we propose the following composite likelihood ratio
test statistic $G_{\mathrm{CLRT}}^2$ for testing the null hypothesis
(\ref{h0}) defined as
\[
G_{\mathrm{CLRT}}^2=-2\log\frac{\sup_{\mathcal{H}_0} \prod_{j=1}^m
\prod_{t=q+1}^{n_j}P\{\vX_{t-q}^{(j)},\vX
_{t-q+1}^{(j)},\ldots,\vX_{t}^{(j)};
\theta=g(\varphi)\}} {\sup_{\mathcal{H}} \prod_{j=1}^m
\prod_{t=q+1}^{n_j}P(\vX_{t-q}^{(j)},\vX
_{t-q+1}^{(j)},\ldots,\vX_{t}^{(j)};\theta)},
\]
which is the likelihood ratio statistic based on the piled data
treating the data as independent. Because of the dependency of the
data, the null distribution of the statistic $G_{\mathrm{CLRT}}^2$ is not the
standard $\chi^2_r$ distribution.

Denote $\theta_0$ as the true value of $\theta$ under the null $H_0$
and $\varphi_0$ as the corresponding~$\phi$, that is, $\theta_0 =
g(\varphi_0)$. Define the $k$-dimensional column vector
\[
\vZ_t = \frac{\partial}{\partial\theta} \log P(\vX_{t-q}, \vX
_{t-q+1},\ldots, \vX_{t};\theta)\bigg|_{\theta= \theta_0},
\]
which is a stationary
ergodic sequence. Let $\vZ_t^{(j)}$ be the corresponding sample
value for $j=1,\ldots, m$. Assuming the standard regularity
conditions on the joint probability density function
$P(\vX_{t-q}^{(j)}, \vX_{t-q+1}^{(j)},\ldots, \vX_{t}^{(j)};\theta)$
as commonly assumed for the likelihood ratio test statistics
[\citet{Cox1979}], Theorem 2 in the supplementary material
[\citet{DenGenLi13}], for
the case of a long time
series and few replications, shows that the asymptotic distribution
of the CLRT statistic $G_{\mathrm{CLRT}}^2$ follows a mixture of $\chi^2$
distributions, not a simple $\chi^2$ distribution as with
conventional LRT statistics. Theorem 3 shows a similar result
when $m\rightarrow\infty$ and $n_j = n$ for all $j$, that is, a
short time
series with many replications. The exact statements of both
theorems and their proofs can be found in the supplementary material
[\citet{DenGenLi13}].

\subsection{Conditional independence tests for the Gaussian DAG models}
In this section we consider the Gaussian DAGs for continuous
random variables. In the MTS setting, we assume that
$(\vX_{t-q}',\ldots, \vX_{t-1}',\vX_{t}')'$ follows a multivariate
normal distribution $N(0,\Sigma)$ where the DAG determines the local
dependency structures of the variables and therefore
the corresponding covariance matrix $\Sigma$. In our tsPCD-PCD
algorithm, the tests for conditional independence can be written as
$H_0\dvtx X_{t,a} \perp X_{t-l,b} | S_{t,q}$, where $\{ X_{t,a},
X_{t-l,b}, S_{t,q} \}$ $\subseteq A=\{\vX_{t-q}',\ldots,
\vX_{t-1}',\vX_{t}'\}$ with $0 \leq l \leq q$ and $a,b \in\{
1,\ldots,p \}$, and $S_{t,q}$ is a separator set. The
corresponding CLRT statistic can be written as
\[
G_{\mathrm{CLRT}}^2=-2 \log\frac{\sup_{\mathcal{H}_0}\prod_{j=1}^m \prod
_{t=q+1}^{n_j} P(X_{t,a}^{(j)},
X_{t-l,b}^{(j)},
S_{t,q}^{(j)};\theta=g(\phi))}{\sup_{\mathcal{H}}\prod_{j=1}^m
\prod_{t=q+1}^{n_j} P(X_{t,a}^{(j)}, X_{t-l,b}^{(j)},
S_{t,q}^{(j)};\theta)}.
\]

Corollary 1 in the supplementary material [\citet{DenGenLi13}]
shows that under the
Gaussian DAG model and the null hypothesis $H_0\dvtx\break X_{t,a}
\perp X_{t-l,b} | S_{t,q}$, ${G_{\mathrm{CLRT}}^2}/{\hat{\lambda}}$ follows a
$\chi^2_1$ distribution, where the exact expression of
$\hat{\lambda}$ can also be found.

%

\section{Simulation studies}\label{simu}
In this section we evaluate the performance of the tsPCD-PCD algorithm
in learning
the local directed graphical structure. We consider the dynamic graph
[Figure \ref{Fig4}(a)] used in Section \ref{example} with time
series data of lag 1. For each simulation, we simulate a training data
set from a joint Gaussian distribution using a structural equation
model of recursive
linear regressions derived from the assumed DAG structure with residual
variances of
$1$. The regression coefficients are randomly generated uniformly from
$(-0.6,-0.2)\cup(0.2,0.6)$ or from $(-0.6,-0.4)\cup(0.4,0.6)$. We
consider models with different sample sizes
and both data from one long time series and data from multiple short
time series.
These parameters were chosen to give reasonable signal-to-noise ratios
for the given sample sizes.
We repeat the simulations 100 times and obtain the average values of the
performance scores for each case.

Similar to the example in Section \ref{example}, our goal is to obtain the
local graph around the node $T=20$ at the depth $d=1$. In the true
graph, node 20 connects with many other nodes and has the largest
degree.

\subsection{Validity of the CLRT statistic}
We first demonstrate that the CLRT statistic $G_{\mathrm{CLRT}}^2$ does not
follow the standard
$\chi^2$-distribution and instead it follows a rescaled $\chi^2$ distribution.
We consider two null hypotheses
$H_0^{\prime}\dvtx\break  X_{t,24} \perp X_{t,2}$ and $H_0^{\prime\prime}\dvtx X_{t,4} \perp
X_{t,1}|\{X_{t-1,2},X_{t,2}\}$ in the model of one single time series of
length $n=500$ and the regression coefficients being generated
uniformly from $(-0.6,-0.4)\cup(0.4,0.6)$. For $H_0^{\prime}$ and
%
%
\begin{figure}

\includegraphics{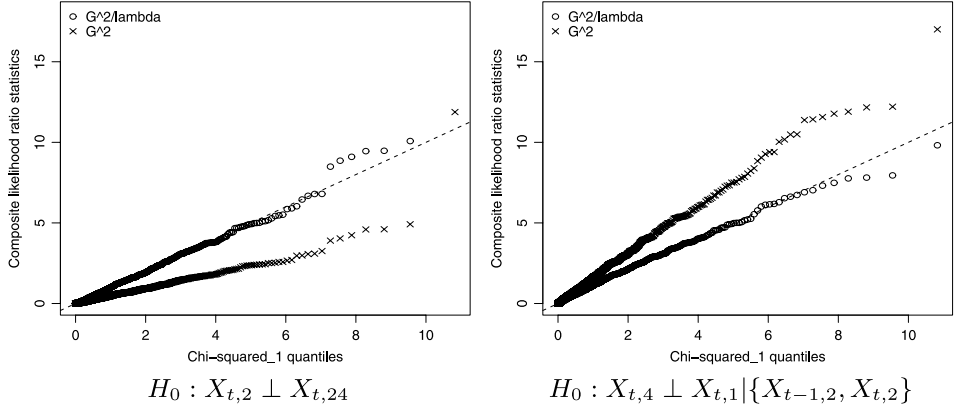}

\caption{Distribution of the CLRT statistic under two null hypotheses
based on the
simulated data, demonstrating that the CLRT statistic follows a
rescaled $\chi^2_1$
distribution.}\label{qqplotfig}\vspace*{6pt}
\end{figure}
$H_0^{\prime\prime}$, $\lambda$ is around 1.8 and 1.2, respectively. We can
see from the Q--Q plots over 1000 simulations in Figure
\ref{qqplotfig} that the CLRT statistic $G_{\mathrm{CLRT}}^2$ greatly
deviates from $\chi^2_1$, while the rescaled statistic
${G_1^2}/{\hat{\lambda}} \rightarrow\chi^2_1$, which indicates that
the asymptotic result of the CLRT statistic holds.

\subsection{Performance in recovering the local directed graph}
To evaluate the performance of tsPCD-PCD in recovering the local
directed graphical structure, for each of three subsets, parents
(Pa), children (Ch) and all depth $1$ variables (PC), we calculate
the score pair (precision, recall), where Precision and Recall or
sensitivity are defined as
\[
\mbox{precision}=\frac{\mathrm{no.}\ \mathrm{of}\ \mathrm{true}\ \mathrm{positives}}
{\mathrm{no.}\ \mathrm{of}\ \mathrm{edges}\ \mathrm{identified}},\qquad \mbox{recall}
=\frac{\mathrm{no.}\ \mathrm{of}\ \mathrm{true}\ \mathrm{positives}}
{\mathrm{no.}\ \mathrm{of}\ \mathrm{true}\ \mathrm{edges}}.
\]

Table \ref{tsPCD} shows the
precision and recall results for four different models when
different test statistics with significance level $\alpha=0.01$ for
the conditional independence tests are used. Overall, we observe
that the set PC(20) can be identified very well in all four models.
In addition, multiple time series resulted in similar results as
%
%
\begin{table}
\def\arraystretch{0.9}
\caption{Precision and recall of the local DAG graph around node
$20$ in the DAG shown in Figure~\protect\ref{Fig4}\textup{(a)} based on 100
replications using the tsPCD-PCD or PCD-PCD algorithm. The composite
likelihood ratio tests are used for testing the conditional
independence with $\alpha=0.01$ with/without including the
rescaling factor $\hat{\lambda}$ to adjust for the dependency of
the observations}\label{tsPCD}
\begin{tabular*}{\tablewidth}{@{\extracolsep{\fill}}lccccccc@{}}
\hline
& & \multicolumn{3}{c}{\textbf{Precision}}
& \multicolumn{3}{c@{}}{\textbf{Recall}} \\[-4pt]
& & \multicolumn{3}{c}{\hrulefill}
& \multicolumn{3}{c@{}}{\hrulefill} \\
\textbf{Statistic} &\textbf{Method}& \textbf{Pa} & \textbf{Ch} & \textbf{PC}
& \textbf{Pa} & \textbf{Ch} & \textbf{PC} \\
\hline
&
\multicolumn{6}{c@{}}{$n=500, M=1$, $\beta\in(-0.6,-0.2)\cup(0.2,0.6)$}
\\[4pt]
$G_{\mathrm{CLRT}}^2$ 
&tsPCD-PCD & 0.50 & 0.82 & 0.97 & 0.44 & 0.47 & 0.73 \\
&PCD-PCD & 0.37 & 0.60 & 1.00 & 0.16 & 0.64 & 0.70 \\
$G_{\mathrm{CLRT}}^2/\hat{\lambda}$ 
&tsPCD-PCD&0.55 & 0.82 & 0.98 & 0.41 & 0.59 & 0.72 \\
&PCD-PCD & 0.35 & 0.58 & 1.00 & 0.16 & 0.61 & 0.71 \\[4pt]
&
\multicolumn{6}{c@{}}{$n=10,M=50$, $\beta\in(-0.6,-0.2)\cup(0.2,0.6)$}
\\[4pt]
$G_{\mathrm{CLRT}}^2$ 
& tsPCD-PCD&0.50 & 0.83 & 0.97 & 0.43 & 0.47 & 0.72 \\
&PCD-PCD& 0.41 & 0.62 & 1.00 & 0.16 & 0.69 & 0.70\\
$G_{\mathrm{CLRT}}^2/\hat{\lambda}$ 
&tsPCD-PCD & 0.53 & 0.83 & 0.97 & 0.42 & 0.55 & 0.71\\
&PCD-PCD& 0.41 & 0.63 & 1.00 & 0.19 & 0.64 & 0.70\\[4pt]
&
\multicolumn{6}{c@{}}{$n=500,M=1,\beta\in(-0.6,-0.4)\cup(0.4,0.6)$}
\\[4pt]
$G_{\mathrm{CLRT}}^2$ 
& tsPCD-PCD&0.42 & 0.91 & 0.90 & 0.27 & 0.64 & 0.60 \\
& PCD-PCD& 0.26 & 0.69 & 1.00 & 0.10 & 0.60 & 0.59 \\
$G_{\mathrm{CLRT}}^2/\hat{\lambda}$ 
&tsPCD-PCD&0.84 & 0.95 & 0.99 & 0.32 & 0.87 & 0.61 \\
&PCD-PCD& 0.32 & 0.68 & 1.00 & 0.14 & 0.56 & 0.59 \\[4pt]
& \multicolumn{6}{c@{}}{$n=1000,M=1,\beta\in(-0.6,-0.2)\cup(0.2,0.6)$}
\\[4pt]
$G_{\mathrm{CLRT}}^2$ 
&tsPCD-PCD&0.35 & 0.82 & 0.82 & 0.25 & 0.70 & 0.64 \\
& PCD-PCD& 0.29 & 0.67 & 0.99 & 0.12 & 0.63 & 0.63 \\
$G_{\mathrm{CLRT}}^2/\hat{\lambda}$ 
& tsPCD-PCD& 0.94 & 0.88 &1.00 & 0.33 & 0.95 & 0.66 \\
& PCD-PCD & 0.30 & 0.62 & 1.00 & 0.13 & 0.59 & 0.64 \\
\hline
\end{tabular*}
\end{table}
single time series when the total number of observations are
comparable. Second, the CLRTs using the correct null distribution
(i.e., mixture of $\chi^2$-distributions) gave better performances
than that using the wrong null distributions in both precision and
recall, especially when the average cross-time correlations are high
or when the sample sizes are large. Almost identical results are
observed when the significance level is set to $\alpha=0.005$.

\subsection{Improved performance when time order is used}
We finally demonstrate that by using the time order information to
orient the edges,
we can substantially increase both the precision and recall rates for
the parents
and children sets. Table \ref{tsPCD} also shows the precision and
recall results from the
standard PCD-PCD algorithm that ignores the time series nature of the
data for the
same sets of models. We observe clear decreases in both
precisions and recalls for the parents and children sets of a node when
the time
order information is ignored, while the selections of all depth 1
variables are
comparable. The results clearly indicate the tsPCD-PCD algorithm that utilizes
the time order to orient the edges can lead to better identification of
the parents
and children nodes of a target variable.


\section{Application to a real data set}\label{real}
The central event in the generation of a cellular immune response to
stimulants is
the activation of T-cells. T-cell activation is initiated
by the interaction between the T-cell receptor complex and the
antigenic peptide presented on the surface
of the cells. Such an activation triggers a network of proteins,
kinases, phosphatases and
adaptor proteins that lead to gene transcription events in the nucleus,
including
transcription of a number of transcription factors such as
c-Fos, c-myc, c-jun and activation of early genes such as
interleukins
(e.g., IL2, IL3R etc.). These genes in turn induce the expression of a
number of
effector genes. Days after the activation event, various adhesion
molecules begin to be expressed. It is therefore very important to
understand the causal relationships among the genes involved in T-cell
activation.

\citet{Rangel2004} measured gene expression levels of $p=58$ genes
that are related to T-cell activation over $n=10$ time points (0, 2, 4,
6, 8, 18, 24, 32, 48, 72 hours) after treating the T-cells with
ionomycin. Expression data over $m=44$ biological replications were
obtained. Following \citet{Rangel2004,Rau2010}, we log-transformed
the expression data and performed the quantile normalization
[\citet {Bolstad2002}] to ensure that all 44 replicates have a
similar underlying distribution of gene expression. The normalized
expression data and gene descriptions can be found in the R package
GeneNet of \citet{Strimmer3} on CRAN [\citet{RDCT2011}].

\begin{figure}[t!]

\includegraphics{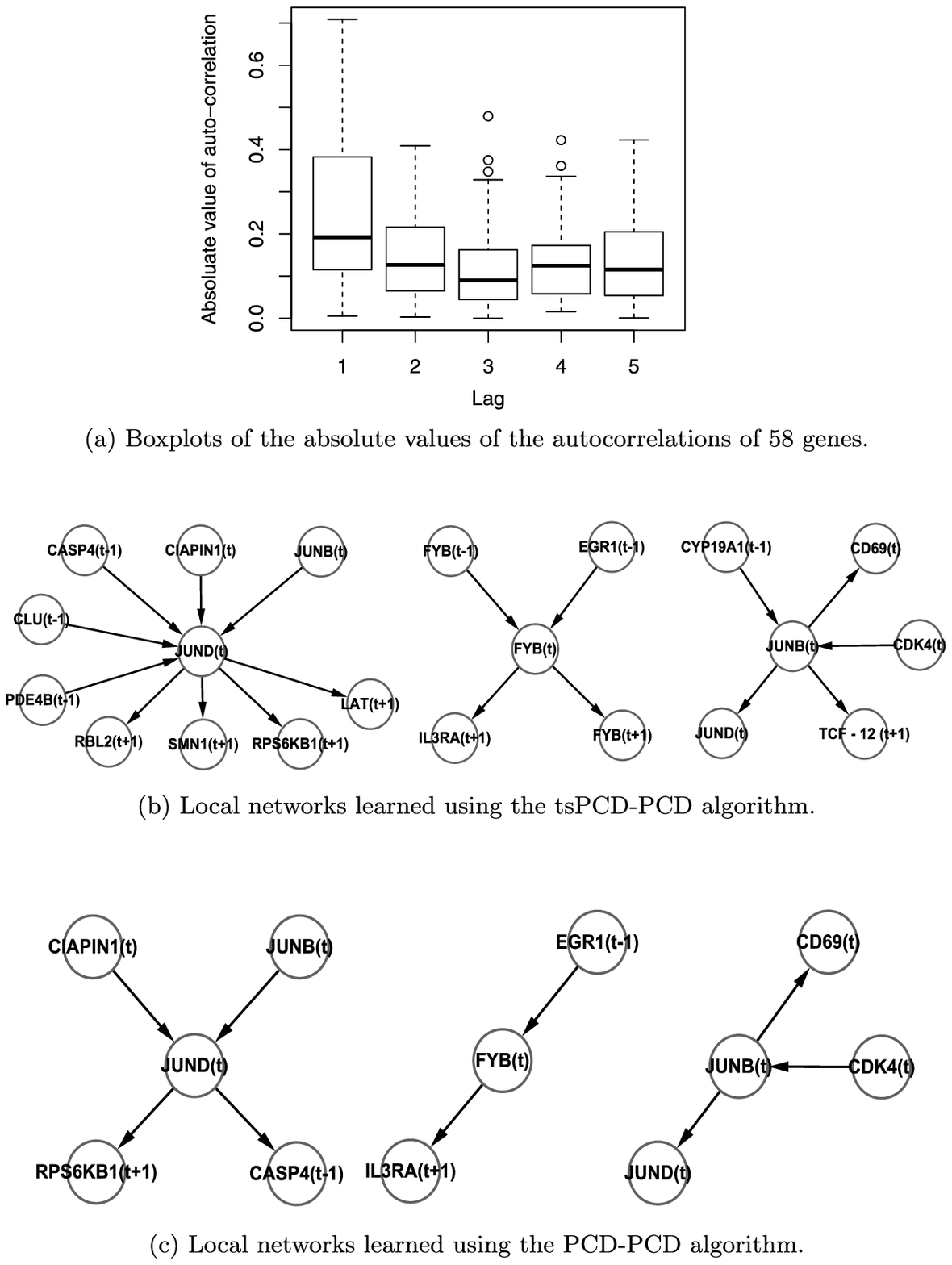}

\caption{Analysis of real data set: boxplots and local structures
learned around genes JUND, FYB and JUNB
based on time course gene expression data using the significant level
$\alpha=0.01$ and $\mbox{depth}=1$.}\label{FigRD3}
\end{figure}

Instead of learning the whole gene regulatory network among these 58
genes related
to Jurket T-cell activation, we focused on learning the local directed
graphs of three
important genes [\citet{Rangel2004,Rau2010}], including the
transcription factors JunB,
JunD and FYB genes using the proposed tsPCD-PCD algorithm.
JunB and c-Jun, along with JunD and Fos group proteins (c-fos,
FosB, Fra1 and Fra2), comprise the core members of the activator
protein 1 (AP1)
family of transcription factors. Since the time series we have are very
short with $n=10$, following \citet{Rau2010} and \citet
{Rangel2004}, we assume
a lag $q=1$ under a Gaussian model to learn local networks around these
three genes
separately. Choosing $q=1$ is partially justified by the boxplots of
the auto-correlations for different lag sizes shown in Figure \ref
{FigRD3}(a). The auto-correlations are small for lags greater than 1.
We used the significance level of $\alpha=0.01$ for all the conditional
independence tests based on the CLRTs or LRTs.


%

Figure \ref{FigRD3}(b) shows the local neighbors identified by the
tsPCD-PCD method for each of
the three genes.
We first observe that genes JunB and JunD
have more neighbors than the FYB gene, showing the importance of these
two genes in
T-cell activation [\citet{Boise1993}]. Out of the five genes that are
found to
regulate expression of JunD (CIAPIN, CLU, CASP, PDE4B and JunB), three
are related to
apoptosis, including the cytokine-induced inhibitor of apoptosis
(CIAPIN) and the
clusterin (CLU) genes that are associated with the clearance of
cellular debris and
apoptosis, and apoptosis-related cycteine peptide (CASP). This is very
interesting
since the T-cells were stimulated by the calcium ionophore ionomucin,
which is
known to induce apoptosis [\citet{Miyake1999}] by activating the
apoptosis-related
genes. These apoptosis events then lead to activation of JunD, which in term
regulates genes of survival of motor neuron 1 (SMN1), ribosomal protein
kinase S6
(RPS6Kb1), retinoblastoma like protein 2 (RBL2) and linker gene for
activation of
T-cells (LAT), all at the next time point. This observation agrees with
the known fact that JunD mediates
survival signaling to mount an appropriate biological response to a specific
challenge [\citet{Lamb2003}]. Among these genes, LAT plays an important
role in the
activation, homeostasis and regulatory function of T cells [\citet
{Shen2010}] and
RBL2 is a key regulator of entry into cell division and survival. The ribosomal
protein S6 kinase (RPS6K1) is a central regulator of protein synthesis
and of cell
proliferation, differentiation and survival [\citet{Han2006}]. This
example shows the effectiveness
of our methods in identifying the upstream regulators and downstream
regulated genes of JunD during
T-cell activation. This is in contrast to the global network identified
by \citet{Rau2010} using a state-space model, where only Caspase-4 was
identified as the upstream regulator of JunD.

JunB is a cell cycle-regulated transcription factor that is involved in
the regulation of a broad spectrum of
cellular functions, including the expression of leukocyte
early activation antigen CD69 [\citet{Castellanos1997}]. It also
interacts with JunD. tsPCD-PCD identifies these two genes as the
downstream targets of JunB.
Cyclin-dependent kinases (CDK4) and their targets have been primarily
associated with
regulation of cell-cycle progression. \citet{Vanden2011} recently
identified
JunB as a newly recognized CDK substrate, supporting the fact that CDK4
is a upstream regulator of JunB. Our method also identifies CYP19A as
another potential regulator of JunB.

Finally, the early-growth response 1 (EGR-1) transcription regulator
was first
identified as an immediate-early response gene transcriptionally
activated by
mitogenic stimulation [\citet{Sukhatme1988}]. It regulates the FYN
binding protein
(FYB), which is an important adaptor molecule in the T-cell receptor signaling
machinery that in turn influences the expression of the interleukin 3
receptor, $\alpha$~(IL3RA). These results agree with the current literature. In contrast,
in work by \citet{Rangel2004},
FYB was found to occupy
a crucial position in the graph and was involved in the highest number of
outward connections. However, our local DAG does not support this
conclusion. It is also interesting to see that
FYB tends to self-regulate over time as shown by the identified edges
FYB($t-1)$ $\rightarrow$ FYB($t$) $\rightarrow$ FYB($t+1$).

As a comparison, we also applied the PCD-PCD algorithm that ignores the time
series nature of the data and shows the resulting local networks in Figure
\ref{FigRD3}(b). Ignoring the time dependency leads to less
well-connected graphs
compared with the ones obtained by tsPCD-PCD. For example, important apoptosis
related JunD regulators CLU and CASP were not identified by the PCD-PCD
algorithm.
This indicates that our algorithm is able to extract additional
information about the
interactions among the investigated genes based on the time-course gene
expression
data.

\section{Discussion}\label{disc}
Motivated by analysis of time course gene expression data with
replicates, we have
developed a learning algorithm to identify the neighboring nodes of a
given variable
by extending the PCD-PCD algorithm of \citet{Zhou2010} in order to
effectively utilize the
time-order in orienting the edges. Like many constraint-based methods,
our algorithm depends on
valid tests for conditional independence. To account for the dependency
among the observations in time series data,
we developed composite likelihood ratio tests that provide valid tests
for conditional independence for general parametric DAG models,
including the log-linear and the Gaussian DAG models. While there are other
alternative tests for continuous variables without assuming
a functional form between the variables as well as the
data distributions such as the kernel-based tests [\citet
{kernel}], it
is not clear how to extend these tests to the multivariate time series
data in the context of learning the DAGs.

Theorem \ref{thm1} of this paper shows that the proposed tsPCD-PCD
algorithm can recover the true local DAG (up to a Markov equivalent
class) if the DAG is faithful to the joint probability distribution and
if all the conditional independence conditions can be correctly
checked with the data. Therefore, the power of discovering the true
local network depends on the power and type 1 error of the conditional
independence tests. However, since many
such conditional independence tests are
performed in the algorithm,
it does not seem to be possible to develop a general framework to
determine the sample sizes needed in order to recover the true DAG with
a high probability. This is an interesting topic for future research.

In our analysis of the gene expression data, the tsPCD-PCD
algorithm has identified several important regulatory relationships
among the genes in T-cell activation pathways; many agree
with the our current knowledge on T-cell activation. The fact that many
edges identified by our method can be substantiated by the literature
shows its effectiveness in identifying biologically useful information
on gene regulation. As a necessary simplification, we assume that DAG
structure is time-invariant.
When data are collected over many time points and
with many replications such as those measured at single-cell levels, it
is possible to extend our methods for
studying time-dependent DAG structures that can reflect the dynamic
changes of the DAG structures.

\section*{Acknowledgments}

We thank You Zhou and Changzhang Wang for sharing their source codes
for the PCD-PCD algorithm. We thank Professor Karen Kafadar, the Associate
Editor and the reviewers for very helpful comments and suggestions that
have led to great improvement in the presentation.

\begin{supplement}
\stitle{Supplemental materials for ``Learning local directed acyclic
graphs based on multivariate time series data''}
\slink[doi]{10.1214/13-AOAS635SUPP} 
\sdatatype{.pdf}
\sfilename{aoas635\_supp.pdf}
\sdescription{The online supplemental materials include detailed
statements and their proofs of Theorems \ref{thm1}--3, Lemma 2 and
Corollary 1. Proof of Theorem \ref{thm1} provides a detailed explanation of the
steps of parts I and II of the tsPCD-PCD algorithm.}
\end{supplement}

%
%

\printaddresses

\end{document}